\documentclass[prd,twocolumn,showpacs]{revtex4}\begin{document}\preprint{}
\title{Relativistic diffusion  }
\author{ Z. Haba\\
Institute of Theoretical Physics, University of Wroclaw,\\ 50-204
Wroclaw, Plac Maxa Borna 9, Poland}

\email{zhab@ift.uni.wroc.pl}\date{\today}
\begin{abstract}
We discuss a  relativistic diffusion in the proper time in  an
approach of Schay and Dudley. We derive (Langevin) stochastic
differential equations in various coordinates. We show that in
some coordinates the stochastic differential equations become
linear. We obtain momentum probability distribution in an explicit
form. We discuss a relativistic particle diffusing in an external
electromagnetic field. We solve the Langevin equations in the case
of parallel electric and magnetic fields. We derive a kinetic
equation for the evolution of the probability distribution.  We
discuss  drag terms leading to an equilibrium distribution. The
relativistic analog of the Ornstein-Uhlenbeck process is not
unique.
 We show that  if the drag comes from a diffusion
approximation to the master equation then its form is strongly
restricted. The drag leading to the Tsallis equilibrium
distribution satisfies this restriction  whereas the one of the
J\"uttner  distribution does not.
 We show that any function of the relativistic energy can be the equilibrium distribution
 for a particle in a static electric field.
 A preliminary study of the time evolution with friction is
 presented. It is shown that the problem is equivalent to quantum
 mechanics of a particle moving on a hyperboloid with a potential
 determined by the drag.
 A relation to
diffusions appearing in heavy ion collisions is briefly discussed.
 \end{abstract}  \maketitle
 \section{Introduction}
 There were many attempts to generalize the diffusion in
 a way respecting relativistic invariance and causality
 (\cite{schay}\cite{dudley}\cite{hakim}\cite{deb}\cite{dunkel}; for
a review and further references see \cite{deb2} \cite{talkner}).
In this paper we develop the approach initiated by Schay
\cite{schay} and Dudley \cite{dudley}. The diffusion process which
should be considered as a relativistic analog of the Brownian
motion is uniqely defined by the requirement that this is the
diffusion whose four momentum stays on the mass-shell. We discuss
Ito stochastic differential equation \cite{ikeda} defined as a
perturbation of the relativistic dynamics on the phase space. As
an example we consider the motion in an electromagnetic field
\cite{ll} (its generalization, a motion in the Yang-Mills field
\cite{kelly}\cite{elze}, could be treated in a similar way).
Relativistic stochastic dynamics preserving the particle's mass
has no normalizable Lorentz invariant equilibrium measure. We
discuss drags which lead to an equilibrium probability measure for
a large time but violate the Lorentz invariance. A stochastic
process with such a drag would be an analog of the
Ornstein-Uhlenbeck process. Covariant drags describing a
relativistic particle in a medium moving with a velocity $V$ are
discussed in \cite{deb}. In such a case the Lorentz transformation
of the friction is compensated  by the transformation of the
velocity $V$.

A diffusion process can be considered as a relativistic
approximation to more complex many particle processes. In
particular, a motion of a heavy particle in an environment of a
gas of light particles, described in a Markovian approximation by
the master equation \cite{kampen}, could be approximated by a
diffusion process \cite{landau}. Such an approximation is applied
in a description of the quark-gluon plasma
\cite{hwa}\cite{svetitsky}\cite{sinha}\cite{raf}\cite{blaizot}\cite{elze},
for an electron in a  background cosmic radiation
\cite{ens}\cite{itoh} or a particle moving in a fluctuating metric
\cite{sorkin}. We show that an equilibrium distribution consistent
with the diffusion approximation to the master equation is
severely restricted. The Tsallis \cite{tsallis} distribution
satisfies this restriction whereas the J\"uttner distribution
\cite{jut} and quantum distributions do not. We discuss the form
of the relativistic diffusion equation and compare it to the
relativistic diffusions discussed in heavy ion collisions.

We explicitly work out solutions to the stochastic dynamics and
its transition function in various coordinate systems. We
determine the momentum distribution of a relativistic particle in
external electromagnetic fields. We discuss the time evolution
with a friction leading to an equilibrium. We show that such a
dynamics is equivalent to an imaginary time evolution of a quantum
mechanical particle moving on the hyperboloid in a potential
determined by the drag. We expect that standard quantum mechanics
methods can be applied for a detailed approximation of the
diffusive evolution. The explicit formulas may be useful for a
comparison of theoretical predictions with experimental results of
ultra-relativistic collisions when a gas of relativistic particles
is formed.
\section{Relativistic dynamics}
 We are interested in random perturbations of the dynamics of relativistic
 particles of mass $m$. On the Minkowski space the dynamics of
 a relativistic particle is described by the equations \cite{ll}
 \begin{equation}
 \frac{dx^{\mu}}{d\tau}=\frac{1}{m}p^{\mu}
 \end{equation}
 \begin{equation}
 \frac{dp^{\mu}}{d\tau}=K^{\mu}
 \end{equation}
 where $\mu=0,1,2,3$ and $K^{\mu}$ is a force.
The four-momentum  $p(\tau)$ of a  relativistic particle defines
the mass by the relation
\begin{equation}
p^{2}=p_{0}(\tau)^{2}-p_{1}(\tau)^{2}-p_{2}(\tau)^{2}-p_{3}(\tau)^{2}=m^{2}c^{2}
\end{equation}Eq.(3) (together with the positivity of energy)
says that the momenta stay on the upper half ${\cal H}_{+}$
(defined by $p_{0}\geq 0$) of the four-dimensional hyperboloid
${\cal H}$. If eq.(3) is satisfied then from eq.(1) it follows
that $\tau$ has the meaning of the proper time. If eq.(3) is to be
true then the force $K^{\mu}$ must satisfy the subsidiary
condition
\begin{equation} p_{\mu}K^{\mu}=0\end{equation}
(we use the convention of a summation over repeated indices). We
add a random force $k_{\mu}d\tau=dp_{\mu}^{H} $ to eq.(2) writing
it in the form
 \begin{equation}
dp_{\mu}=K_{\mu}d\tau+dp_{\mu}^{H}
 \end{equation}

 The diffusion $p^{H}(\tau)$ on the hyperboloid  ${\cal H}_{+}$ is uniquely defined. It  is generated by the
Laplace-Beltrami operator on ${\cal H}$
\begin{equation}
\triangle_{H}=\frac{1}{\sqrt{g}}\partial_{j}g^{jk}\sqrt{g}\partial_{k}
\end{equation}
Here, $g=\det(g_{jk}) $ and $g_{jk}$ is the metric on ${\cal H}$.
If we define the expectation value over the sample paths (starting
from $(x,p)$) of the diffusion process
$\phi_{\tau}(x,p)=E[\phi(x(\tau),p(\tau))]$ (we denote the
expectation values by $E[...]$) then $\phi_{\tau}$ satisfies the
diffusion equation
\begin{equation}
\partial_{\tau}\phi_{\tau}= \Big(\frac{p^{\mu}}{m}
 \frac{\partial}{\partial x^{\mu}}+K^{\mu}
 \frac{\partial}{\partial p^{\mu}}+\frac{\gamma^{2}}{2}\triangle_{H}\Big)\phi_{\tau}
\end{equation}
with the initial condition $\phi$. $\gamma^{2}$ has the meaning of
a diffusion constant. We write it in the form
\begin{equation} \gamma=mc \kappa \end{equation} Then,
$\kappa^{-2}$ has the dimension of time.

\section{ Coordinates on ${\cal H}$}
A proper choice of coordinates may be useful for a solution of
differential equations. The momenta (3) ${\bf p}$ (with
$p_{0}=\sqrt{{\bf p}^{2}+m^{2}c^{2}}$) could be used as
coordinates
 on ${\cal H}_{+}$. Then, the metric tensor can be obtained from
 the embedding of ${\cal H}$  in $R^{4}$. We obtain
(expressing $dp_{0}$ by
 $dp_{k}$ from eq.(3)) \begin{equation} g_{jk}=\delta^{jk}-p_{0}^{-2}p^{j}p^{k}
\end{equation}
Then, $g=m^{2}c^{2}p_{0}^{-2}$ and
\begin{equation}
g^{jk}=\delta^{jk}+(mc)^{-2}p^{j}p^{k}
\end{equation}
Hence,
\begin{equation}
\triangle_{H}=\partial_{1}^{2}+\partial_{2}^{2}+\partial_{3}^{2}+(mc)^{-2}p^{j}p^{k}\partial_{j}\partial_{k}
+(mc)^{-2}3p^{k}\partial_{k}
\end{equation}
where $k=1,2,3$ and $\partial_{j}=\frac{\partial}{\partial p^{j}}$.
We prefer another choice of coordinates $(p_{+},p_{a})$ (where
$a=1,2$)
\begin{displaymath}
p_{+}=p_{0}+p_{3}
\end{displaymath}
In such a case $g_{++}=p_{+}^{-2}(p_{a}p_{a}+m^{2}c^{2})$,
$g_{+a}=-p_{a}p_{+}^{-1}$ and $g_{ab}=\delta_{ab}$. Then, $
g=m^{2}c^{2}p_{+}^{-2}$, $g^{++}=(mc)^{-2}p_{+}^{2}$,
 $g^{+a}=(mc)^{-2}p_{+}p_{a}$,
$g^{aa}=1+(mc)^{-2}p_{a}^{2}$, $g^{12}=(mc)^{-2}p_{1}p_{2}$.
Hence,
\begin{equation}\begin{array}{l}
\triangle_{H}=\partial_{1}^{2}+\partial_{2}^{2}+(mc)^{-2}p_{+}^{2}\partial_{+}^{2}
+(mc)^{-2}p_{1}^{2}\partial_{1}^{2}\cr+
(mc)^{-2}p_{2}^{2}\partial_{2}^{2}+2(mc)^{-2}p_{+}p_{a}\partial_{a}\partial_{+}
+ 2(mc)^{-2}p_{1}p_{2}\partial_{1}\partial_{2} \cr
+(mc)^{-2}3p_{+}\partial_{+}+(mc)^{-2}3p_{a}\partial_{a}\end{array}\end{equation}
where $\partial_{a}=\frac{\partial}{\partial p_{a}}$. The formula
(12) can be rewritten in a Lorentz invariant form
\begin{equation}
\begin{array}{l}
\triangle_{H}=-3(mc)^{-2}p^{\mu}\partial_{\mu}+(mc)^{-2}(p^{\mu}p^{\nu}-\eta^{\mu\nu}p^{2})\partial_{\mu}\partial_{\nu}
\end{array}\end{equation}
where $\eta^{\mu\nu}$ is the Minkowski metric. In order to derive
eq.(12) from eq.(13) assume that $p^{2}=m^{2}c^{2}$ and the function
$\phi$ in eq.(7) is expressed as a function of $p_{+},p_{1},p_{2}$.

It is instructive to compare ${\bf p}$ with some other widely used
coordinates (a change of coordinates in the relativistic diffusion
equation is also discussed in \cite{deb2}\cite{bal}).
 First, let us consider the analogues of spherical coordinates
\begin{equation}
p_{0}=mc\cosh \alpha
\end{equation}
\begin{displaymath}
p_{1}=mc\sinh\alpha\cos\phi\sin\theta
\end{displaymath}
\begin{displaymath}
p_{2}=mc\sinh\alpha\sin\phi\sin\theta
\end{displaymath}\begin{displaymath}
p_{3}=mc\sinh\alpha\cos\theta
\end{displaymath}
In these coordinates the metric is expressed as
\begin{equation}
ds^{2}=(mc)^{2}(d\alpha^{2}+(\sinh \alpha)^{2}ds_{2}^{2})
\end{equation} where \begin{equation}
 ds_{2}^{2}=d\theta^{2}+(\sin\theta)^{2}d\phi^{2}
 \end{equation}
 is the metric on the sphere $S_{2}$.
 The Laplace-Beltrami operator reads
 \begin{equation}
 (mc)^{2}\triangle_{H}=(\sinh
 \alpha)^{-2}\partial_{\alpha}(\sinh\alpha)^{2}\partial_{\alpha}+
 (\sinh\alpha)^{-2}\triangle_{S_{2}}
\end{equation}
where  $\triangle_{S_{2}}$ is the Laplace-Beltrami operator on the
sphere. Next, let us consider the Poincare coordinates
$(q_{1},q_{2},q_{3})$ on ${\cal H}_{+}$ which are related to
momenta $p$ as follows
\begin{equation}
p_{3}+p_{0}=\frac{mc}{q_{3}}
\end{equation}\begin{displaymath}
p_{3}-p_{0}=-\frac{mc}{q_{3}}(q_{1}^{2}+q_{2}^{2}+q_{3}^{2})
\end{displaymath}\begin{displaymath}
p_{1}=\frac{mc q_{1}}{q_{3}}
\end{displaymath}\begin{displaymath}
p_{2}=\frac{mc q_{2}}{q_{3}}
\end{displaymath}
where $q_{3}\geq 0$. Then, the metric is
\begin{equation}
ds^{2}=(mc)^{2}q_{3}^{-2}
 (dq_{1}^{2}+dq_{2}^{2}+dq_{3}^{2})
\end{equation}
and
\begin{equation}(mc)^{2}\triangle_{H}=q_{3}^{2}(\partial_{1}^{2}+\partial_{2}^{2}+\partial_{3}^{2})
-q_{3}\partial_{3}
\end{equation}
\section{Stochastic equations}
A solution of the  diffusion equation as well as expectation
values of observables can be expressed by an expectation value
over the solution of stochastic equations \cite{ikeda}. We have
discussed  stochastic equations corresponding to the diffusion on
${\cal H}_{+}$ and their solutions in
\cite{habaijmp}\cite{habaprd}. The stochastic equations on ${\cal
H}_{+}$ are also discussed in \cite{deb2}\cite{jan}\cite{bal}. We
solve these equations in the case of the free motion ($K=0$) using
the Poincare coordinates or the light-cone coordinates. In the
Poincare coordinates the diffusion process is a solution of the
linear stochastic differential equations
\begin{equation}
dq_{a}=\kappa q_{3}db_{a}
\end{equation}
$a=1,2$.
\begin{displaymath}
dq_{3}=-\frac{\kappa^{2}}{2}q_{3}d\tau+\kappa
q_{3}db_{3}=-\kappa^{2}q_{3}d\tau+\kappa q_{3}\circ db_{3}
\end{displaymath}
where Stratonovitch differentials are denoted by a circle and the
Ito stochastic differentials without the circle (the notation is
the same as in \cite{ikeda}). The Brownian motion appearing on the
rhs of  eqs.(21) is defined as the Gaussian process with the
covariance
\begin{equation}
E[b_{a}(\tau)b_{c}(s)]=\delta_{ac}min(\tau,s) \end{equation}
 The solution of eqs.(21) is
\begin{equation}
q_{3}(\tau)=\exp(-\kappa^{2}\tau+\kappa b_{3}(\tau))q_{3}
\end{equation}
and
\begin{equation}
q_{a}(\tau)=q_{a}+\kappa\int_{0}^{\tau}q_{3}(s)db_{a}(s)
\end{equation}
for $a=1,2$. The solution could be applied for a calculation of
correlation functions and the transition function. The transition
function $P_{\tau}$ of the diffusion is a solution
 of the  equation\begin{equation}
\partial_{\tau}P=\frac{\gamma^{2}}{2}\triangle_{H}P
\end{equation}
with the initial condition
$P_{0}(q,q^{\prime})=g^{-\frac{1}{2}}\delta(q-q^{\prime})$. We
have \cite{ikeda} (we calculated the transition function from the
solution of the stochastic equations in \cite{habaprd})
\begin{equation}\begin{array}{l} P_{\tau}(\sigma) =(2\pi
\kappa^{2}
\tau)^{-\frac{3}{2}}\sigma(\sinh\sigma)^{-1}\exp(-\frac{\kappa^{2}\tau}{2}-\frac{\sigma^{2}}{2\kappa^{2}\tau})
\end{array}\end{equation}
where the geodesic distance $\sigma$ in the Poincare coordinates can
be expressed in the form

\begin{equation}\cosh \sigma=1+(2q_{3}q_{3}^{\prime})^{-1}(( q_{1}-
q_{1}^{\prime})^{2}+(q_{2}-q_{2}^{\prime})^{2}+(q_{3}-q_{3}^{\prime})^{2})\end{equation}
  Using eqs.(18) we can derive the differentials
$dp_{\mu}^{H}$. In the light-cone coordinates
\begin{equation}
p_{\pm}=p_{0}\pm p_{3}
\end{equation}
we have
\begin{equation}
dp_{+}=\kappa^{2}p_{+}d\tau+\kappa p_{+}\circ
db_{+}=\frac{3\kappa^{2}}{2}p_{+}d\tau+\kappa p_{+} db_{+}
\end{equation} (here we denoted $b_{3}$ by $b_{+}$)\begin{equation}
dp_{a}=\frac{3}{2}\kappa^{2}p_{a}d\tau+\kappa p_{a} db_{+}+\gamma
db_{a}
\end{equation}
where $a=1,2$

$p_{-}$ can be obtained from the formula
\begin{equation}\begin{array}{l}
p_{-}=(m^{2}c^{2}+p_{a}p_{a})p_{+}^{-1}\end{array}
\end{equation}
Then \begin{displaymath}\begin{array}{l} dp_{-}=\kappa^{2}
(2p_{-}- \frac{3m^{2}c^{2}}{p_{+}})d\tau \cr
+\kappa
(p_{-}-\frac{2m^{2}c^{2}}{p_{+}})\circ db_{+} +
\frac{2\gamma}{p_{+}}p_{a}\circ db_{a}
\end{array}\end{displaymath}Let
\begin{equation}
\phi_{\tau}(p)=E[\phi(p(\tau))]
\end{equation}where $p(\tau)$ is the solution of  stochastic
equations (29)-(30) with the initial condition $p$.
Then,\begin{equation}
\partial_{\tau}\phi_{\tau}=\frac{\gamma^{2}}{2}\triangle_{ H}\phi_{\tau}
\end{equation}
where $\triangle_{H}$ is defined in eq.(12).

The spatial momenta which are useful for a physical interpretation
of the relativistic diffusion lead to a non-linear (Ito) Langevin
equation
\begin{equation}
dp^{j}=\frac{3}{2}\kappa^{2}p^{j}d\tau+ e^{j}_{n}({\bf p})db^{n}
\end{equation}
where \begin{equation} g^{jk}=e^{j}_{n}e^{k}_{n}
\end{equation}
with
\begin{equation}
e^{j}_{n}=\delta^{jn}+(p_{0}-mc)(mc)^{-1}{\bf p}^{-2}p^{j}p^{n}
\end{equation}
These equations have been derived earlier in \cite{deb2}. We can
solve these equations by means of a change of coordinates (18)
applying the solutions (23)-(24) or (29)-(30).
 {\section{Phase space evolution in an electromagnetic
field} The evolution of coordinates can be obtained as an integral
over the proper time
\begin{equation}
x_{\mu}(\tau)=x_{\mu}+\frac{1}{m}\int_{0}^{\tau}p_{\mu}(s)ds
\end{equation}
In an electromagnetic field the momentum satisfies the equation
\begin{equation}
dp_{\mu}=\frac{e}{mc}F_{\mu\nu}p^{\nu}d\tau+dp^{H}_{\mu}
\end{equation}
Here, by $p^{H}$ we denote the diffusion on the hyperboloid
defined in eqs.(29)-(30). In general, we obtain non-linear
stochastic differential equations from eq.(38) (because eq.(31)
for $p_{-}$ is non-linear in momentum). Eq.(38) is a linear
equation if the equation for $p_{+}$ does not involve $p_{-}$ on
the rhs. The only case which leads to linear stochastic
differential equations describes constant parallel electric and
magnetic fields (or  a special case when one of them is zero). Let
\begin{equation}
\alpha=\frac{e}{mc}
\end{equation}
Assume that the only components of $F$ are $F_{12}=B$ and
$F_{30}=E$. In such a case eqs.(38) read
\begin{equation}
dp_{1}=\alpha Bp_{2}d\tau+\frac{3}{2}\kappa^{2}p_{1}d\tau+\kappa
p_{1} db_{+}+\gamma db_{1}
\end{equation}
\begin{equation}
dp_{2}=-\alpha Bp_{1}d\tau+\frac{3}{2}\kappa^{2}p_{2}d\tau+\kappa
p_{2} db_{+}+\gamma db_{2}
\end{equation}
\begin{equation}
dp_{+}=\alpha Ep_{+}d\tau +\kappa^{2}p_{+}d\tau+\kappa p_{+}\circ
db_{+}
\end{equation}
It is clear that the linear equations (40)-(42) can explicitly be
solved. The solution of eq.(42) is an elementary
function\begin{equation} p_{+}(\tau)=p_{+}\exp(\alpha
E\tau+\kappa^{2}\tau+\kappa b_{+}(\tau))
\end{equation}
The environment of electromagnetic waves or (in a quantized form)
photons can be another source of diffusion. Let in eq.(2)
$K_{\mu}=(F_{\mu\nu}+Q_{\mu \nu})p^{\nu}$ where $Q$ is a Gaussian
electromagnetic field (depending only on the proper time) with the
covariance
\begin{equation}
E[Q_{\mu\nu}(\tau)Q_{\sigma\rho}(\tau^{\prime})]=(\eta_{\mu\sigma}\eta_{\nu\rho}-
\eta_{\mu\rho}\eta_{\nu\sigma})\delta(\tau-\tau^{\prime})
\end{equation}
Let $p(\tau)$ be the solution of eq.(2) with an external
(deterministic) electromagnetic field $F$ and a random (or
quantum) electromagnetic field $Q$. Then,
$\phi_{\tau}(p)=E[\phi(p(\tau;p))]$ is the solution of eq.(7).

In order to derive  the non-relativistic limit we assume that
\begin{equation}
p_{+}=mc=const
\end{equation}
in the stochastic equations (29)-(30).Then,
\begin{displaymath}
\partial_{+}=\partial_{3}
\end{displaymath}
in the diffusion equation (33).  In the non-relativistic limit
eqs.(38) for a constant electromagnetic field become linear. The
solution is expressed by the Ornstein-Uhlenbeck process
\cite{kurs}.
\section{The momentum distribution}

We are interested in a distribution of momenta of particles coming
out from a gas formed after heavy ion collisions. For this purpose
we express the transition function (26) by the momenta. The
relativistic invariant formula reads
\begin{equation} \cosh\sigma=\frac{1}{2}(mc)^{-2}pp^{\prime}
\end{equation}
In terms of the $(p_{+},p_{1},p_{2})$ coordinates we have
\begin{equation}
\begin{array}{l}
2\cosh\sigma\equiv 2a=m^{-2}c^{-2}p_{+}^{-1} p_{+}^{\prime -1}\cr
\Big( (p_{1}p_{+}^{\prime }- p_{1}^{\prime}p_{+})^{2} \cr
+(p_{2}p_{+}^{\prime }- p_{2}^{\prime}p_{+})^{2}\cr +
m^{2}c^{2}p_{+}^{\prime 2}+ m^{2}c^{2}p_{+}^{2}\Big)
\end{array}\end{equation}
As a function of $a$ the geodesic distance $\sigma$ has the form
\begin{equation}
\sigma=\ln(a+\sqrt{a^{2}-1})
\end{equation}
The time evolution in the momentum coordinates is
 \begin{equation}
 \phi_{\tau}(p)\equiv T_{\tau}\phi(p)=\int d\mu(
 p^{\prime})P_{\tau}(p,p^{\prime})\phi(p^{\prime})
 \end{equation}
 where $d\mu=d^{3}p p_{0}^{-1}mc$ is the relativistic invariant
 volume measure $\mu$. We express eq.(49) in various coordinate systems
 inserting the transition function (26) in eq.(49) with proper volume elements $d\mu$.
  In the coordinates (14) the Riemannian volume element is
\begin{equation}
d\mu=(mc)^{3}d\alpha d\theta d\phi\sinh^{2}\alpha\sin\theta
 \end{equation}
  In the Poincare
coordinates
\begin{equation}
d\mu=(mc)^{3}dq_{1}dq_{2}dq_{3}q_{3}^{-3}
\end{equation}

From the  invariance under Lorentz transformations $\Lambda$
\begin{equation}
\cosh\sigma (p,p^{\prime})=\cosh\sigma(\Lambda p,\Lambda
p^{\prime}).
\end{equation}
We can choose
\begin{equation}
\Lambda p=(p_{0},0,0,p_{3})
\end{equation}
Let us define the rapidity $y$ by
\begin{equation}
p_{0}\pm p_{3}=m_{T}c\exp(\pm y)
\end{equation}
where
\begin{equation}
m_{T}^{2}c^{2}=m^{2}c^{2}+p_{T}^{2}=m^{2}c^{2}+p_{1}^{2}+p_{2}^{2}
\end{equation}
The rapidity transforms in a simple way under the Lorentz boost
(with the velocity $v$ ) in the $(0,3)$ plane
\begin{equation}
\tilde{ y}=y+\frac{v}{c}
\end{equation}
 Then, in the frame where $p_{T}=0$
\begin{equation}
\sigma(p,p^{\prime})=y-y^{\prime}
\end{equation}
The rapidity is also closely related to the variable $\alpha$ in
the coordinates (14). We have
\begin{displaymath}
\cosh\sigma=\cosh\alpha\cosh\alpha^{\prime}-\sinh\alpha\sinh\alpha^{\prime}\cos\sigma_{2}
\end{displaymath}
where $\sigma_{2}$ is the geodesic distance on the unit sphere.
Hence, in  the Lorentz frame (53) if $\sigma_{2}=0$ then
$\sigma=y-y^{\prime}=\alpha-\alpha^{\prime}$.

From eqs.(29) and (43) we can see that the process $p_{+}(\tau)$ is
an exponential of a Gaussian process. Hence, its probability
distribution should be the log-normal distribution. We could
calculate it from the general formula (using the transition function
(26))\begin{equation}\begin{array}{l} \phi_{\tau}(p_{+})=\int
dp_{1}^{\prime}dp_{2}^{\prime}dp_{+}^{\prime}p_{+}^{\prime
-1}P_{\tau}(p,p^{\prime})\phi(p_{+}^{\prime})\cr \equiv \int
dp_{+}^{\prime}P^{(+)}_{\tau}(p_{+},p_{+}^{\prime})\phi(p_{+}^{\prime})
\end{array}\end{equation}
However, it is easier to derive it directly from the solution (43).
So, for the diffusion in the electric field (40)-(42) we obtain
\begin{equation}\begin{array}{l}
P^{(+)}_{\tau}(p_{+},p_{+}^{\prime})_{E}=(2\pi\kappa^{2}\tau)^{-\frac{1}{2}}
(p_{+}^{\prime})^{\kappa^{-2}\alpha E}p_{+}^{-1-\kappa^{-2}\alpha
E}\cr \exp\Big(-\frac{\tau}{2\kappa^{2}}(\alpha
E+\kappa^{2})^{2}-\frac{1}{2\kappa^{2}\tau}(\ln(\frac{p_{+}^{\prime}}{p_{+}}))^{2}\Big)
\end{array}
\end{equation}

\section{The equilibrium distribution}
Let us define the time evolution of an expectation value of an
observable $\phi$  in a state $\rho$ (a measure on the phase
space) by\begin{equation} \langle \phi\rangle^{\tau}_{\rho}=\int
d\rho_{\tau}\phi \equiv \int d\rho \phi_{\tau}\end{equation} We
say that a measure $\nu$ is the invariant measure for the
diffusion process (see \cite{ikeda}) if the expectation value in
eq.(60) is time-independent, i.e.
\begin{equation}
\int d\nu(p,x)\phi_{\tau}(p,x)=const
\end{equation} Assume that (in general) the diffusion
equation reads
\begin{displaymath}
\partial_{\tau}\phi_{\tau}={\cal G}\phi_{\tau}\end{displaymath}
where
\begin{equation}
{\cal G}=\frac{\gamma^{2}}{2}\triangle_{H}+Y
\end{equation}and
\begin{equation}Y=R^{j}\frac{\partial}{\partial
p^{j}}+\frac{p^{\mu}}{m}\frac{\partial}{\partial
x^{\mu}}\end{equation}is the generator of the deterministic flow
in the coordinates (10)
or\begin{equation}Y=R_{a}\frac{\partial}{\partial
p_{a}}+R_{-}\frac{\partial}{\partial
p_{+}}+\frac{p^{\mu}}{m}\frac{\partial}{\partial
x^{\mu}}\end{equation}in the coordinates (12). Let us write
\begin{equation}
d\rho_{\tau}=dxd^{3}p\Phi_{\tau} \end{equation} Then, from eq.(60)

\begin{equation}
\partial_{\tau}\Phi_{\tau}={\cal G}^{*}\Phi_{\tau}
\end{equation}
where in the coordinates (10)
\begin{equation}\begin{array}{l}
{\cal
G}^{*}=\frac{\gamma^{2}}{2}\triangle_{H}^{*}-\frac{\partial}{\partial
p_{j}}R_{j}-\frac{p^{\mu}}{m}\frac{\partial}{\partial x^{\mu}}
\end{array}\end{equation}
and
\begin{equation}
\triangle_{H}^{*}=\partial_{j}g^{jk}\sqrt{g}\partial_{k}\frac{1}{\sqrt{g}}
\end{equation}
Let us write the invariant measure in the form
\begin{equation}
d\nu= d^{3}pdx\sqrt{g}\Phi_{R}\equiv d^{3}pdx\Phi_{0}\Phi_{R}
\end{equation}
Here, $d^{3}p\Phi_{0}=d^{3}p\sqrt{g}$ is the (not normalizable;
$g$ is calculated below eq.(9)) equilibrium measure for
$\triangle_{H}$, i.e.
\begin{equation}
\triangle_{H}^{*}\Phi_{0}=0
\end{equation}

 Then, the invariant measure $\nu$ for the diffusion (62) is determined by the
solution $\Phi_{R}$ of the equation (obtained by differentiating
eq.(61) over $\tau$)
\begin{equation}
{\cal G}^{*}\Phi_{0}\Phi_{R}=0
\end{equation}
or \begin{equation} \tilde{{\cal G}}\Phi_{R}=0
\end{equation}where
\begin{equation}\begin{array}{l}
\tilde{{\cal
G}}=\frac{\gamma^{2}}{2}\triangle_{H}-p_{0}(\frac{\partial}{\partial
p_{j}}R_{j}+\frac{\partial}{\partial
x^{\mu}}\frac{p^{\mu}}{m})p_{0}^{-1}
\end{array}\end{equation}
 It can easily be seen that if the (non-zero) limit of
$\rho_{\tau}$ (as $\tau\rightarrow \infty$) exists then
\begin{equation}\lim_{\tau\rightarrow\infty}d\rho_{\tau}=dxd^{3}pp_{0}^{-1}\Phi_{R}
\end{equation}
(in the weak sense of the convergence of measures). We can express
eq.(72) as an evolution equation in time $x^{0}$
\begin{equation}\begin{array}{l}
\partial_{0}\Phi_{R}
=\frac{\kappa^{2}}{2}mp_{0}^{-1}\triangle_{H}\Phi_{R}
-m\Big(\frac{\partial}{\partial
p_{j}}R_{j}+p_{j}\frac{\partial}{\partial
x^{j}}\Big)p_{0}^{-1}\Phi_{R}\end{array}
\end{equation}
If $R^{j}$ does not depend on $x^{0}$ then eqs.(66) and (72) may
have the same time-independent solutions determining the static
equilibrium distribution $\Phi_{E}$. In general, eq.(75) is the
transport equation for $\Phi_{R}$. When $x^{0}\rightarrow \infty$
then $\Phi_{R}$ tends to the $x^{0}$-independent equilibrium
distribution $\Phi_{E}$
\begin{equation}
\lim_{x_{0}\rightarrow\infty}\Phi_{R}=\Phi_{E} \end{equation}
solving both eqs.(66) and (75).

Eq.(72) can be considered as an equation for the drag  if the
equilibrium measure $\Phi_{E}$ is fixed. It is not possible to
obtain the equilibrium measure which is normalizable, Lorentz
invariant and at the same time concentrated on the mass-shell
($p^{2}=m^{2}c^{2}$). We give up the explicit Lorentz invariance.
If we still require the rotation invariance then it is useful to
work in the spherical coordinates (14). In these coordinates
eq.(72)  for $\Phi_{E}$ reads (we restrict ourselves to the
momentum dependence of $\Phi_{E}$)
\begin{equation}
\frac{\gamma^{2}}{2}\triangle_{H}\Phi_{E}=p_{0}\partial_{\alpha}(\omega\Phi_{E})
\end{equation}
where
\begin{equation}
Y=\omega(\alpha)\frac{\partial}{\partial \alpha}
\end{equation}
of eq.(62) in the coordinates (14) has only one component
$\omega$. From eq.(77) we obtain
\begin{equation}
\omega=\frac{1}{2}\gamma^{2}\partial_{\alpha}\ln\Phi_{E}
\end{equation}
If
\begin{equation}
\Phi_{E}=\exp(-\beta c p_{0})=\exp(-\beta mc^{2}\cosh\alpha)
\end{equation}
then
\begin{equation}
\omega=-\frac{1}{2}\gamma^{2}mc^{2}\beta\sinh\alpha
\end{equation}
For the Bose-Einstein distribution
\begin{equation}
\Phi_{E}=\Big(\exp(\beta mc^{2}\cosh\alpha)-1\Big)^{-1}
\end{equation}
we have \begin{equation}
\omega=-\frac{1}{2}\gamma^{2}mc^{2}\beta\sinh\alpha
\Big(1-\exp(-\beta mc^{2}\cosh\alpha)\Big)^{-1}\end{equation} The
drifts can be inserted (after a change of coordinates) into the
diffusion equation (7) or the stochastic equations (29)-(30) in
order to determine the diffusive dynamics.

The equilibrium distribution (80) determines the diffusion
generator in spherical coordinates
\begin{equation}
\begin{array}{l}
{\cal
G}=\frac{\gamma^{2}}{2m^{2}c^{2}}p^{0}u^{-2}\frac{\partial}{\partial
u}p^{0}u^{2}\frac{\partial}{\partial u}
+\frac{\gamma^{2}}{2u^{2}}\triangle_{S_{2}}-\frac{1}{2}\kappa^{2}\beta
c p^{0}u\frac{\partial}{\partial u}
\end{array}\end{equation}
where $u=\vert{\bf p}\vert$ and $p^{0}=\sqrt{m^{2}c^{2}+u^{2}}$.
If the drift $R^{j}$ for $\Phi_{E}$ (80) is derived from eq.(72)
in the  ${\bf p}$ coordinates (10) then we obtain
\begin{equation}
{\cal
G}=\frac{\gamma^{2}}{2}\triangle_{H}-\frac{1}{2}\kappa^{2}\beta c
p_{0}p^{j}\frac{\partial}{\partial p^{j}}
\end{equation}

\section{Diffusion equation as an approximation to master
equation} The diffusion equation (7) could be considered as an
approximation to the dynamics of a heavy particle embedded in a
gas of light particles. The kinetic equation describing the flow
conservation  under a Markovian scattering process reads
\cite{kampen}\cite{landau}(here $t=\frac{x_{0}}{c}$)
\begin{equation}\begin{array}{l}
(\partial_{t}+cp^{j}p_{0}^{-1}\frac{\partial}{\partial
x^{j}})\rho({\bf p},x)\cr =\tilde{\kappa}^{2}\int
d^{3}k\Big(w({\bf p}+{\bf k},{\bf k})\rho({\bf p}+{\bf
k},x)-w({\bf p},{\bf k})\rho({\bf p},x)\Big)\end{array}
\end{equation}
 where $\int d^{3}k w({\bf p},{\bf k})=1$ and
 $\tilde{\kappa}^{2} w({\bf p},{\bf k})$
  is the probability that in the unit time the momentum ${\bf p}$
  of the heavy particle is changed to
${\bf p}-{\bf k}$ through scattering on light particles (we could
 assume $\tilde{\kappa}=\kappa$ but this is not necessary). The
diffusion equation can be obtained by means of the Taylor
expansion in ${\bf k}$ \cite{landau}. In such a case the diffusion
coefficients can be calculated using the formulas
\begin{equation}
C^{j}=\tilde{\kappa}^{2}\int d^{3}k w({\bf p},{\bf k})k^{j}\equiv
\tilde{\kappa}^{2}\langle k^{j}\rangle\end{equation}
\begin{equation}\frac{1}{2}D^{ij}=\frac{1}{2}\tilde{\kappa}^{2}\int d^{3}k w({\bf
p},{\bf
k})k^{i}k^{j}\end{equation}Then,\begin{equation}\begin{array}{l}
\frac{1}{2}D^{ij}=\frac{1}{2}\tilde{\kappa}^{2}\int d^{3}k w({\bf
p},{\bf k})(k^{i}-\langle k^{i}\rangle)(k^{j}-\langle
k^{j}\rangle)\cr+\frac{1}{2}\tilde{\kappa}^{2}\langle
k^{i}\rangle\langle k^{j}\rangle \equiv
\frac{1}{2}M^{ij}+\frac{1}{2}\tilde{\kappa}^{-2}C^{i}C^{j}\end{array}
\end{equation}
It follows that if $w({\bf p},{\bf k})\geq 0$  then the matrix
\begin{equation} M^{ij}=D^{ij}-\tilde{\kappa}^{-2}C^{i}C^{j} \end{equation} must be
positive definite.

The drift and diffusion coefficients have the simplest meaning in
the coordinates (10). In these coordinates, comparing eqs.(7),
(11),(75) and (86) we obtain the diffusion equation for
$\Phi_{R}=\rho$ with
\begin{equation}
C^{j}=\frac{mc}{p_{0}}(\frac{3}{2}\kappa^{2}p^{j}+R^{j})
\end{equation}From eqs.(10) and (90)
\begin{equation}M^{ij}=\frac{mc}{p_{0}}g^{ij}-\tilde{\kappa}^{-2}C^{i}C^{j}
\end{equation}
must be a positive definite matrix ( the metric tensor $g^{ij}$ is
defined in eq.(10)). Let us assume that
\begin{equation}
\Phi_{E}=\exp(f(\beta p_{0}c))
\end{equation}
In such a case from eq.(72) we obtain
\begin{equation}
R_{j}=\frac{1}{2}p_{j}p_{0}\beta\kappa^{2}cf^{\prime}(\beta p_{0}
c)
\end{equation}
Hence,
\begin{equation}
\begin{array}{l}
M_{jk}=\gamma^{2}(\delta_{jk}-p_{j}p_{k}{\bf
p}^{-2})\frac{mc}{p_{0}} \cr
+p_{j}p_{k}\gamma^{2}(mc)^{-1}p_{0}^{-2}{\bf
p}^{-2}\Big(p_{0}^{3}-mc\kappa^{2}\tilde{\kappa}^{-2}{\bf
p}^{2}(\frac{3}{2}+\frac{1}{2}p_{0}\beta cf^{\prime})^{2}\Big)
\end{array}\end{equation}
For J\"uttner \cite{jut} as well as Bose-Einstein equilibrium
distributions the longitudinal term in eq.(95) becomes negative at
large momenta. Hence, at large energies the diffusion equation
could not be a good approximation to the master equation. However,
for the Tsallis distribution \cite{tsallis}
\begin{equation}
\Phi_{E}(x)=(1+(q-1)x)^{-\frac{1}{q-1}}
\end{equation}we have
\begin{equation}
R_{j}=-\frac{1}{2}p_{j}p_{0}\beta\kappa^{2}c(1+(q-1)\beta
cp_{0})^{-1}
\end{equation}
$xf^{\prime}(x)$ in eq.(94) is a bounded function. As a
consequence the longitudinal term in eq.(95) is positive definite.
In such a case we can apply the diffusion equation as an
approximation of the master equation at high momenta as well. The
Tsallis distribution appears in many models ranging from
turbulence \cite{beck} to heavy ion collisions
\cite{raf}\cite{wilk}\cite{beck2}.

We return to the proper time evolution of the relativistic
diffusion in an electromagnetic field of sec.5. We restrict
ourselves to the static case without a magnetic field. Then, the
electromagnetic potential is $A=(A_{0},0,0,0)$. We notice that
${\cal G}$ in eq.(62) consists of two parts : the diffusion in
momentum and the drag which is a sum of the dynamical piece and a
dissipative one. It is easy to see that in the equation  (72) for
the equilibrium measure the dissipative and dynamical parts must
separately vanish (this is a version of the
fluctuation-dissipation theorem). Hence,
\begin{equation}
d\nu=d^{3}xd^{3}pp_{0}^{-1}\exp(f(\beta(cp_{0}+eA_{0}))
)\end{equation} is the equilibrium measure in a time-independent
electric field ${\bf E}=-\nabla A_{0}$. The  drag term of eq.(62)
reads
\begin{equation} Y=\frac{1}{2}\kappa^{2}\beta c p_{0}p^{j}f^{\prime}
(\beta p_{0}c+\beta eA_{0})\frac{\partial}{\partial
p^{j}}-\frac{e}{c}\partial^{j}A_{0} \frac{\partial}{\partial
p^{j}}+\frac{p^{\mu}}{m}\frac{\partial}{\partial x^{\mu}}
\end{equation}
If we suggest that the master equation (86) is an equation in the
proper time $\tau$ (instead of $t=\frac{x_{0}}{c}$) then there
would be no damping factor $\frac{mc}{p_{0}}$ in the drift (91).
In such a case the longitudinal term would become negative for
large momenta. As a consequence  the relativistic diffusion
equation (7) could not be a consistent approximation at large
energies to the proper time master equation .
\section{Evolution with friction}
 In the case of time-independent drags (as discussed in sec.7) the
search of equilibrium can equivalently be treated as a study of
either proper time evolution of $x_{0}$ -independent densities
$\Phi$ or $x_{0}$-independent solutions od the transport equation
(75). Without friction the proper time evolution is expressed
explicitly by eqs.(49) and (26). With friction we do not expect
explicit solutions. We must rely on approximate methods or
computer simulations. There is no normalizable invariant measure
for the relativistic diffusion generated by $\triangle_{H}$
because the process is fast growing as can be seen from the
solution (43) (without the electric field)
\begin{equation}
E[p_{+}^{2}(\tau)]=p_{+}^{2}\exp(4\kappa^{2}\tau)
\end{equation}
The friction is damping the time evolution. Without the noise term
the proper time evolution is determined by the solution of the
equation \begin{equation} \frac{d{\bf p}}{d\tau}={\bf R}({\bf p})
\end{equation} generated by the flow $Y$ (63). The evolution
in $x_{0}$ is defined by the drag of eq.(75)
\begin{equation}
 \frac{d{\bf p}}{dx_{0}}=\frac{m}{p_{0}}{\bf R}({\bf p})
\end{equation}
As an example, for the generator (85) (without the diffusion) we
have
\begin{displaymath}
{\bf p}(x_{0})=\exp(-\kappa^{2}\beta mc x_{0}){\bf p}
\end{displaymath}
The evolution in proper time (101) is also expressed by an
elementary function which  has the asymptotic behavior (coinciding
with the one in the time $\frac{x^{0}}{c}$)
\begin{equation}
\vert {\bf p}(\tau)\vert \simeq \vert {\bf p}(\tau)\vert
\exp(-\kappa^{2}mc^{2}\beta \tau))
\end{equation} The dynamical systems (101)-(102) have a trivial limit
 when time tends to infinity. The diffusive spreading makes it
non-trivial. The evolution with friction is described by the
semigroup $\exp(\tau(\frac{1}{2}\triangle_{H}+Y))$. A rough
approximation of this evolution as a product
$\exp(\tau\frac{1}{2}\triangle_{H})\exp(\tau Y)$ can be expressed
as a diffusion acting on the deterministic flow (101). Such an
approximation is reliable only for a small time.

        Before we propose an exact method to approach the time evolution
        with friction let us explain it
in the well-known case of the  Ornstein-Uhlenbeck (OU) process.
The evolution is generated by
\begin{equation}
{\cal G}_{OU}=\frac{1}{2}\frac{d^{2}}{d\xi^{2}} -\omega
\xi\frac{d}{d\xi}
\end{equation}
Applying the invariant measure for the OU process we can transform
the generator (104) into the Hamiltonian of the harmonic
oscillator
\begin{equation}
H_{osc}=
-\frac{1}{2}\frac{d^{2}}{d\xi^{2}}+\frac{\omega^{2}}{2}\xi^{2}-\frac{\omega}{2}
=-\exp(-\frac{\omega}{2}\xi^{2}){\cal
G}_{OU}\exp(\frac{\omega}{2}\xi^{2} )\end{equation} We apply the
method to the diffusion with friction. We have for the generator
(62) (with the friction (94))
\begin{equation}
-\exp(f){\cal G}\exp(-f)=-\frac{\gamma^{2}}{2}\triangle_{H}+V
\end{equation}
where
\begin{equation}
V=\frac{\kappa^{2}}{2}(c\beta)^{2}{\bf p}^{2}(f^{\prime
2}+f^{\prime\prime})+\frac{3}{2}\kappa^{2}\beta cp_{0}f^{\prime}
\end{equation}
We have obtained a Hamiltonian for a particle moving on the
hyperboloid (3) in a potential V. We could apply either
Hamiltonian methods of quantum mechanics  to this model
\cite{vanholten} or functional integration. In the latter case,
let $\phi=\exp(-f)\psi$, then we have the Feynman-Kac formula
\begin{equation}
\begin{array}{l}
\phi_{\tau}({\bf p},x)=\exp(-f)E[\exp(-\int_{0}^{\tau}V({\bf
p}_{s}^{H})ds)\psi({\bf p}_{\tau}^{H},x_{\tau})]
\end{array}\end{equation} where ${\bf p}_{s}^{H}$ is the stochastic
process (34)-(36) on the hyperboloid (3) starting at the point
${\bf p}$. Clearly, the solution (108) can also be expressed in
the form
\begin{displaymath}
\phi_{\tau}({\bf p},x)=E[\phi({\bf p}_{\tau},x_{\tau})]
\end{displaymath}
where ${\bf p}_{\tau}$ is the solution of the
equation
\begin{equation} d{\bf
p}_{\tau}={\bf R}d\tau+d{\bf p}_{\tau}^{H}
\end{equation}
with the initial condition ${\bf p}$.

Let us note that if $f$ is determined by the relativistic Boltzman
equilibrium distribution (80) (or Bose-Einstein or J\"uttner) then
the potential $V$ in eq.(107) is growing quadratically. Such a
quadratic growth can substantially change the large time behavior
of the solution of the diffusion equation. In the case of the
Tsallis distribution the potential $V$ is bounded. The expansion
in $V$ of eq.(108) (which coincides with the Dyson expansion of
quantum mechanics) is convergent for arbitrarily large time. If
the evolution of the momentum distribution tends to the Tsallis
distribution then this should also be visible from the solution
(108) even far from the equilibrium.

\section{Discussion }

It is  expected that the relativistic diffusion will appear in
relativistic models of plasma. Some calculations based on the
master equation (86) have been performed already in the eighties
\cite{hwa}\cite{svetitsky}(for a recent review see \cite{rapp2}).
The scattering probabilities can be calculated in quantum field
theory. Another approach applies the Wigner function \cite{zachar}
for a description of the particle phase space evolution
 \cite{elze}\cite{blaizot}\cite{gul}.

The approach based on the master equation leads to the diffusion
equation (75) for the probability density. We may write the
diffusion part of eq.(75) (no friction) as
\begin{equation}
\triangle_{H}^{*}=\partial_{i}\partial_{j}D_{ij}-\partial_{i}A_{i}
\end{equation}
where
\begin{equation}
D_{ij}=\gamma^{2}(\delta_{ij}-p_{i}p_{j}{\bf
p}^{-2})\frac{mc}{p_{0}} + \gamma^{2}p_{i}p_{j}{\bf
p}^{-2}\frac{p_{0}}{mc}
\end{equation}
and
\begin{equation}
A_{i}=\gamma^{2}\frac{3}{2}\frac{1}{p_{0}mc}p_{i}
\end{equation}
The total drift (with the friction) is \begin{equation}
C_{i}=\kappa^{2}p_{i}\Big(\frac{3}{2}+\frac{1}{2}\beta
p_{0}cf^{\prime}(\beta p_{0}c)\Big)\frac{mc}{p_{0}}
\end{equation}
The authors \cite{svetitsky}\cite{raf}\cite{rapp}\cite{rapp2}
write a general form of the diffusion equation without specifying
the diffusion coefficients. It follows from our work that  the
diffusion coefficients $D^{ij}$ are defined by the relativistic
invariance. Then, the functions $B_{\parallel}$ and $B_{\perp}$ in
eq.(11) of \cite{raf} and  (the analogs) $B_{1}$ and $B_{0}$ in
eq.(25) of \cite{rapp2} are uniquely determined. Relativistic
invariance determines also the coefficient $A$ of the drift in
eq.(10) of ref.\cite{raf} as $\frac{3mc}{2p_{0}}$ if friction is
switched off. The drag defining the friction is in one to one
correspondence with the equilibrium measure as discussed already
in \cite{raf}.

We have shown that the equilibrium measures resulting from the
diffusion approximation to the master equation (86) which are
selected by the additivity of entropy (J\"uttner or Bose-Einstein)
do not lead to a diffusion equation which could be valid at
arbitrarily high energies. The stochastic process with the
generator (85) could have been considered as a relativistic
generalization of the Ornstein-Uhlenbeck (OU)(see \cite{deb} for
another definition of the relativistic OU process). However, the
relativistic counterpart of the Maxwell-Boltzman equilibrium
measure does not seem to be restricted to the J\"uttner
distribution ( see the discussion in
\cite{dunkel2}\cite{raf}\cite{rafelski2}\cite{beck2}). For this
reason all the stochastic processes with the drifts (113) could be
considered as  relativistic OU processes (they have the usual OU
non-relativistic limit).

  The transport equation resulting from quantum field
theory has an application to ultra-relativistic collisions (in
particular to heavy ion collisions). The time evolution of
observables (denoted by $\phi$) or the probability density of the
diffusion process (denoted by $\Phi$) has been expressed in this
paper by analytic formulas or by equations which could be solved
numerically. The results are determined by the relativistic
invariance and the form of the invariant measure. In this way the
experimental results concerning the particle momentum distribution
(available from RHIC \cite{RHIC} and future experiments on LHC)
can decide whether the particles coming out from relativistic
collisions can be described as diffusing in a gas of light
particles. A diffusion equation can also be treated as a tool for
a phenomenological description of the scattering data in heavy ion
collisions (see \cite{rapp}\cite{japan}).

 Finally, let us mention that although there is only one diffusion on the hyperboloid (3),
  nevertheless, there are many Markov processes on this
hyperboloid. We have a general Levy-Khintchin representation
formula for processes with independent increments. Among these
processes  we distinguish here the fractional diffusion (with a
stable probability distribution) defined by
\begin{equation}
\partial_{\tau}\phi=(-\triangle_{H})^{\delta}\phi
\end{equation}
where $0<\delta\leq 1$. The soluble case $\delta=\frac{1}{2}$ with
the transition function
\begin{equation}\begin{array}{l}
P_{\tau}(\sigma)=\tau\sqrt{2}\pi^{-\frac{3}{2}}\frac{\sigma}{\sinh
\sigma}(\tau^{2}+\sigma^{2})^{-1}
K_{2}(\sqrt{\tau^{2}+\sigma^{2}})
\end{array}\end{equation}
(where $K_{\nu}$ is the Bessel function of the third kind) shows
characteristic features of  the fractional diffusion.  An
application of the fractional diffusion in heavy ion collisions
has been suggested recently in \cite{hun}.

\end{document}